\documentclass[aps]{revtex4}

\begin{document}
\draft

%
%  Uncomment following two lines and one below for 2 column format.
%@twocolumnfalse\endcsname
%\tightenlines

\preprint{Nisho-07/1}
\title{Spontaneous Breakdown of U(1) symmetry \\
in DLCQ without Zero Mode}
\author{Aiichi Iwazaki}
\address{Department of Physics, Nishogakusha University, Ohi Kashiwa Chiba
  277-8585,\ Japan.} 
\date{Feb. 25, 2007} 
\begin{abstract}
We show that the spontaneous breakdown of U(1) symmetry in a Higgs model can be
described in discretized light cone formulation even by neglecting zero mode.  
We obtain correctly the energy of a ground state with the symmetry breakdown. 
We also show explicitly the presence of a Goldstone mode and
its absence when the U(1) symmetry is gauged. In spite of obtaining the favorable results,
we lose a merit in 
the formulation without zero modes
that a naive Fock vacuum is the true ground state.
\end{abstract}	
\hspace*{0.3cm}
\hspace*{1cm}

\maketitle
%\vskip2pc
%%%%%%%%%%%%%%%%%%%%%%%%%%%%%%%%%%%%%%%%%%%%%%%%%
%\tightenlines

The light cone formulation\cite{lc,cone} of quantum field theories 
is very useful for investigating non-perturbative properties of the
theories since the forms of their Hamiltonians seem apparently simple
and tractable, and a naive Fock vacuum is the true vacuum.
It has recently been exploited extensively for the analysis of gluonic structure
of hadrons in high energy scatterings\cite{cgc}.  
But, in order to analyze non-perturbative properties of ground states such as spontaneous symmetry
breakdown, 
we need a careful treatment\cite{yamawaki} of a zero mode, $p^{+}=0$.
It appears explicitly when we quantize it in a box (
$-L\leq x^{-}\leq L$ ). The mode is not a dynamically
independent degree of freedom. It is expressed in terms of
the other dynamical non zero modes with $p^{+}\neq 0$ through so-called zero mode condition.
Its form is very complicated.
Thus, the forms of the Hamiltonians including the zero mode
are not so simple and tractable 
as we expected.

If we obtain effectively correct
results even by neglecting the zero mode,
% in the limit of $L\to \infty$, 
the light cone formulation becomes very appealing
since Hamiltonian without the zero mode becomes very simple.
In the previous papers\cite{thorn} a theory of real scalar field with
a discrete symmetry
has been analyzed by neglecting the zero modes in 1+1 dimensions. 
It has been shown that a coherent state formed by non zero modes
approximately describes the properties of the vacuum with 
the spontaneous breakdown of the discrete
symmetry.  

In this paper 
we analyze a theory of a complex scalar field with
global U(1) symmetry in space-time 4 dimensions.
We apply discretized light cone quantization ( DLCQ ) neglecting the zero mode. 
We obtain a ground state with the nonzero vacuum expectation value of the field.
It is uniform in transverse directions, but not uniform in longitudinal direction.
This contrasts with the case in the standard formulation with the zero mode,
where the ground state is characterized by spatially uniform vacuum expectation value. 
We show that even if we neglect the zero mode, we can obtain correctly the energy of
the ground state.
Furthermore, we can show explicitly the presence of Goldstone boson
in the ground state with the spontaneous symmetry breakdown. On the other hand,
when the symmetry is local, it can be shown that the Goldstone boson
is eaten by gauge fields.
These results make the DLCQ with neglecting the zero mode
useful and valid for practical applications,
although a merit of a Fock vacuum being
the true vacuum is lost.

We quantize the field in a finite longitudinal region
( $-L\leq x^{-}\leq L$ with $x^{-}=(x^0-x^3)/\sqrt{2}$ ) and in
infinite transverse region ($-\infty<\vec{x}<\infty $ with $\vec{x}=x^1,\,x^2$ ).
We take the periodic boundary condition in the longitudinal direction;
$\phi(x^{-}=-L,\vec{x})=\phi(x^{-}=L,\vec{x})$.
Light cone time is denoted as 
$x^{+}=(x^0+x^3)/\sqrt{2}$.
We omit to write down the dependence on the time coordinate, $x^{+}$ of the field. 
We define the zero mode of the field,
$\phi_{p^{+}=0}=\int^{L}_{-L}dx^{-}\phi(x^{-},\vec{x})/2L$, and
remove it from the field. 
Then, the field without the zero mode\cite{thorn} can be expressed in terms of creation and annihilation
operators,

\begin{equation}
\label{phi}
\phi=\sqrt{\frac{\pi}{L}}\sum_{p^{+}>0,k}\frac{1}{\sqrt{2\pi p^{+}}}
\biggl(a_{p^{+},k}e^{-ip^{+}x^{-}}\phi_k(\vec{x})+b^{\dagger}_{p^{+},k}e^{ip^{+}x^{-}}
\phi^{\dagger}_{k}(\vec{x})\biggr)
\end{equation}
where $p^{+}=n\pi/L>0$ ( $n=1,2,3,,,$ ) is the longitudinal momentum and $k$ represents
quantum number characterizing orthonormal eigenfunctions, $\phi_k(\vec{x})$ in the transverse space;
$\int d\vec{x} \phi^{\dagger}_k(\vec{x})\phi_l(\vec{x})=\delta_{k,l}$
and $\sum_k \phi^{\dagger}_k(\vec{x})\phi_k(\vec{y})=\delta(\vec{x}-\vec{y})$.
For example, we may choose $\phi_k(\vec{x})=e^{-i\vec{k}\vec{x}}/2\pi$. In this case, $k$ denotes 
momentum, $\vec{k}$.
The operators, $a_{p^{+},k}$ and $b_{p^{+},k}$, satisfy the commutation relations,
$[a_{p^{+},k},\,a^{\dagger}_{q^{+},l}]=\delta_{p^{+},q^{+}}\delta_{k,l}$,
$[b_{p^{+},k},\,b^{\dagger}_{q^{+},l}]=\delta_{p^{+},q^{+}}\delta_{k,l}$ and the others vanish. 
Then, equal light cone time commutation relation of the field is given by

\begin{equation}
[\partial_-\phi(x^{-},\vec{x}),\phi^{\dagger}(y^{-},\vec{y})]=-i\delta(\vec{x}-\vec{y})
(\delta(x^{-}-y^{-})-1/2L)
\end{equation}
where the last term in the right hand side of the equation comes from
neglecting the zero mode, $\phi_{p^{+}=0}$ of the field.
We have used the identity, 
$\sum_{p^{+}} e^{ip^{+}x^{-}}=\sum_{n=\mbox{integer}} e^{i\pi nx^{-}/L}=2L\delta(x^{-})$.

Now, we analyze ground states of the following Hamiltonian of the field,

\begin{equation}
\label{H}
H(\phi)=\int^{L}_{-L} dx^{-}d\vec{x}\biggl(|\vec{\partial}\phi(x^{-},\vec{x})|^2-m^2|\phi(x^{-},\vec{x})|^2
+\frac{\lambda}{2}|\phi(x^{-},\vec{x})|^4\biggr),
\end{equation}
which represents a Higgs model 
with the global
U(1) symmetry, $\phi\to e^{i\theta}\phi$.
We have assumed the normal ordering of the operators in the Hamiltonian.

Inserting the field in eq(\ref{phi}) into the first two terms of the Hamiltonian,
we obtain 

\begin{equation}
H_0=-\sum_{p^{+}>0,k}\frac{-\vec{k}^2+m^2}{p^{+}}\biggl( a^{\dagger}_{p^{+},k}a_{p^{+},k}
+b^{\dagger}_{p^{+},k}b_{p^{+},k} \biggr),
\end{equation}
where we have chosen explicitly the states, $\phi_k(\vec{x})=e^{-i\vec{k}\vec{x}}/2\pi$. 
Obviously a naive Fock vacuum, $|0\rangle$ ( $a_{p^{+},k}|0\rangle=0$ for all $p^{+}>0$ and all $k$ )
is energetically unstable. States involving particles of $a$ and $b$ 
with small transverse momentum such as $\vec{k}^2<m^2$, 
have lower energies than that of the naive vacuum. 
This point sharply contrasts
with the standard light cone formulation\cite{cone,yama} without neglecting the zero mode.
The naive vacuum is the real vacuum in the formulation. It is one of the great merits
in the light cone quantization. In our approach we lose the merit but we get
a new merit of obtaining a simple Hamiltonian.

Although the creation of the particles arises in the Fock vacuum for lowering the energy, 
the creation does not proceed endlessly because of the presence of 
the repulsive self-interaction, $|\phi|^4$.
Eventually, a coherent state such as $\langle\phi\rangle=ve^{-ip^{+}_0x^{-}}$
( $p^{+}_0>0$ ) 
is obtained as a ground state of the Hamiltonian, where $v=\sqrt{m^2/\lambda}$ minimizes 
the potential energy, $H(\langle\phi\rangle)=-m^2v^2+\lambda v^4/2$.
Therefore, the U(1) symmetry
is broken spontaneously. We note that the ground state
is a coherent state of the non zero mode with $p^{+}_0\neq 0$, while
it is a coherent state of the zero mode in the standard light cone formulation.
It is interesting that the ground state in the case of the complex
scalar field with U(1) symmetry
possesses only a single longitudinal momentum
contrary to the case of the real scalar field\cite{thorn} with discrete symmetry. In the case,
a ground state has various longitudinal momenta.

It should be noted that
the energy of the ground state in the theory of the complex scalar field is the same as the one 
obtained by using semi-classical approximation in the equal-time quantization. 
This contrasts with the case in the real scalar field, in which case
an approximate
ground state\cite{thorn}
has an energy higher than that of the true vacuum obtained semi-classically.
We should also note that the ground state is not uniform
in the longitudinal direction although it is uniform in the 
transverse directions. It depends only on a single momentum, $p^{+}_0>0$,
which can be taken arbitrarily. 

We point out that
since we neglect the zero mode, the momentum, $\langle P^{+}\rangle$, of the ground state
does not vanish for the finite $L$; $\langle P^{+}\rangle=2L(p^{+}_0)^2v^2$. 
But it vanishes
for the infinite length $L\to \infty$ since $\langle P^{+}\rangle
=2(n_0\pi)^2v^2/L$ where $p^{+}_0=n_0\pi/L$.

We proceed to show the presence of Goldstone bosons in the light cone
formulation without the zero mode.
According to the standard procedure, we divide the field as a classical part, $\phi_c$ and
a quantum part, $\phi_q$ describing quantum fluctuations;
$\phi=\phi_c+\phi_q$. $\phi_c$ is given by $ve^{-ip^{+}_0x^{-}}$
and $\phi_q$ is given in eq(\ref{phi}). Then, the Hamiltonian can be rewritten as

\begin{eqnarray}
H&=&H_c+\int^{L}_{-L} dx^{-}d\vec{x}\biggl(|\vec{\partial}\phi_q(x^{-},\vec{x})|^2
-m^2|\phi_q(x^{-},\vec{x})|^2+\frac{\lambda}{2}\biggl(4|\phi_q(x^{-},\vec{x})\,\phi_c|^2+
\phi_c^{\dagger \,2}\phi_q^2(x^{-},\vec{x})  \nonumber \\
&+&\phi_c^2\phi_q^{\dagger \,2}(x^{-},\vec{x})\biggr)+ \mbox{higher orders of $\phi_q$}\biggr) 
\end{eqnarray}
with $H_c=-\int^{L}_{-L} dx^{-}d\vec{x}\,\,m^4/(2\lambda)$, 
where we have only shown the quadratic terms in $\phi_q$.
It is easy to diagonalize the quadratic terms by decomposing $\phi_q$ 
into the real part, $\phi_1$, and the imaginary part, $\phi_2$,; $\phi_q=(\phi_1+i\phi_2)/\sqrt{2}$. 
We obtain 

\begin{equation}
H=H_c+\int^{L}_{-L} dx^{-}d\vec{x}\,\frac{1}{2}\biggl((\vec{\partial}\delta\phi_1)^2+
(\vec{\partial}\delta\phi_2)^2+2m^2\delta\phi_1^2\biggr),  
\end{equation}
with
\begin{equation}
\delta\phi_1=\cos(p^{+}_0x^{-})\phi_1-\sin(p^{+}_0x^{-})\phi_2, \quad 
\delta\phi_2=\sin(p^{+}_0x^{-})\phi_1+\cos(p^{+}_0x^{-})\phi_2. 
\end{equation} 

These two fields commute with each other; $[\delta\phi_1,\delta\phi_2]=0$.
Obviously, the field, $\delta\phi_2$, represents the Goldstone boson in the vacuum
with the spontaneous breakdown of the U(1) symmetry. 
The field with zero transverse momentum ( $\vec{\partial}\delta\phi_2=0$ )
creates a state degenerate with the ground state.

Now, we proceed to show that the Goldstone boson is absorbed in gauge fields
when the symmetry becomes local. 
We assume that gauge fields couple minimally with the field, $\phi$. 
Then, 
by choosing the light cone gauge, $A^{+}=0$,
we obtain the following Hamiltonian\cite{cone},  

\begin{equation}
\label{Hl}
H_{local}=\frac{1}{2}F_{1,2}^2+
|(\vec{\partial}-ig\vec{A})\phi|^2-m^2|\phi|^2+\frac{\lambda}{2}\,|\phi|^4
+\frac{g^2}{2}\rho\frac{1}{-\partial_{-}^2}\rho,
\end{equation}
with U(1) charge $\rho=i(\phi^{\dagger}\partial_{-}\phi-\partial_{-}\phi^{\dagger}\phi)$,
where $F_{1,2}=\partial_1A_2-\partial_2A_1$ is the field strength of U(1) gauge field, $\vec{A}$,
in the transverse directions.
The last term in $H_{local}$ results from
the choice of the light cone gauge, $A^{+}=0$.
This is similar to the case that 
Coulomb interaction term arises in Hamiltonian when we choose Colomb gauge.

We make a comment on the singular operator, $1/(-\partial_{-}^2)$.
The term with the operator in $H$ arises by solving a constraint condition, $\partial_{-}^2A^{-}=\rho$,
for the field, $A^{-}$.
In order to define the operator being regular, we impose a condition 
that a component, $\rho_{n=0}$ of the charge, $\rho=\sum_{n=\mbox{integer}}\rho_n\,e^{ix^{-}p^{+}}$
( $p^{+}=\pi n/L$ ), 
must vanish.
With this condition the gauge field, $A^{-}$ satisfies the periodic
boundary condition. Thus, by subtracting the component, $\rho_{n=0}$,
from the definition of $\rho$, the operator becomes regular.
The condition implies that the total charge vanishes
in the closed space of $S^{1}$; $Q=\int d\vec{x}\int^{L}_{-L}dx^{-}\rho=2L\int d\vec{x}\rho_{n=0}=0$.

We now rewrite the Hamiltonian only by taking quadratic terms of $\delta\phi_i$
as in the previous way,  

\begin{eqnarray}
&|&(\vec{\partial}-ig\vec{A})\phi\,|^2+\frac{g^2}{2}\rho\frac{1}{-\partial_{-}^2}\rho 
\,\simeq (\vec{\partial}\delta\phi_1)^2+(\vec{\partial}\delta\phi_2)^2 \nonumber \\
&-&2g|\phi_c|\vec{A}\vec{\partial}\phi_2+g^2|\phi_c|^2\vec{A}^2 
+g^2|\phi_c|^2(\partial_{-}\delta\phi_2-2p_{0}\delta\phi_1)\frac{1}{-\partial_{-}^2}
(\partial_{-}\delta\phi_2-2p_{0}\delta\phi_1) \nonumber \\ 
&=&(\vec{\partial}\delta\phi_1)^2+g^2|\phi_c|^2(\vec{A}-|\phi_c|\vec{\partial}\delta\phi_2/g)^2
+g^2|\phi_c|^2(\delta\phi_2')^2,
\end{eqnarray}
with $\delta\phi_2'\equiv\delta\phi_2-2p_{0}\int_{-L}^{x^{-}}dy^{-}\delta\phi_1$,
where we have written down only relevant terms for the Goldstone boson.
The other part of the Hamiltonian does not involve the Goldstone boson, $\delta\phi_2$.

By changing the variable such as $\vec{A'}=\vec{A}-|\phi_c|\vec{\partial}\delta\phi_2/g$, we
can see that the Goldstone boson, $\delta\phi_2$ is absorbed in the gauge fields.
As a result, the transverse components of the gauge fields 
gain the mass term, $g^2|\phi_c|^2\vec{A'}^2$.
The last term, $g^2|\phi_c|^2(\delta\phi_2')^2$, represents
the mass term of the longitudinal 
component of the gauge fields, 
which becomes a dynamical
variable due to the spontaneous symmetry breakdown. It is necessary to
discuss this point in detail for clarifying the mechanism
how the gauge fields gain the mass in the light cone formulation. 

In this way we find in the DLCQ without the zero mode 
that the Goldstone boson is present when the symmetry is global, while
it is absent when the symmetry becomes local.
Therefore, the spontaneous symmetry breakdown can be described properly in
DLCQ even without zero mode.

Finally, we wish to point out a utility of DLCQ without zero mode.
As is well known, an effective potential of color magnetic field, $B$, in QCD
has a non trivial minimum, i.e. $B\neq 0$ in one loop approximation\cite{savvidy}. 
Thus,  
the spontaneous generation of the color magnetic field arises. But,
this non trivial vacuum with the color magnetic field is unstable\cite{unstable,nielsen}.
Gluons under the magnetic field have imaginary energies, $E$,
in equal-time quantization.
For example, the gluons in the lowest Landau level 
have such energies that $E^2=k^2-gB$ where $k$ is a momentum parallel to $B$.
Thus, $E$ can be imaginary for sufficiently small $k$
and Hamiltonian is not Hermitian in the one loop approximation.
Such gluons produce an imaginary part in the effective potential at $B\neq 0$.
It means that the non trivial vacuum with $B\neq 0$ is unstable. 
Thus, we need to find a more stable state\cite{iwa} than the non trivial vacuum.
The analysis has been down previously in a semi-classical approximation\cite{nielsen}.
Analysis at full quantum level has not yet been performed because
an appropriate Hermitian Hamiltonian 
has not been obtained in the equal-time quantization.

In such a circumstance, DLCQ without zero mode can lead to
a tractable and appropriate quantum Hamiltonian in the above analysis of QCD.
This is because light cone energies, $P^{-}$, of gluons under the color
magnetic field are real; $P^{-}=-gB/p^{+}$.
Hence, we can define an Hermitian Hamiltonian.
This is an excellent utility for the application of DLCQ to the analysis of QCD. 
The application has recently been performed\cite{iwazaki} in the analysis 
of dense quark matter.
We have found that the condensation of gluons arises just as in the case of the Higgs model
discussed in this paper.
 
To summarize, by using DLCQ without zero modes we can describe the spontaneous breakdown of U(1)
symmetry in a similar way to that in the standard time-like quantization.
This is a very attractive for the application of the DLCQ 
to the analysis of theories involving imaginary energies, e.g. QCD with color magnetic field, 
because light cone energies can be always defined to be real even in such theories.

\vspace*{2em}
We would like to express thanks
to Prof. O. Morimatsu for useful discussion.
This work was supported by Grants-in-Aid of the Japanese Ministry
of Education, Science, Sports, Culture and Technology (No. 13135218).

%%%%%%%%%%%%%%%%%%%%%%%

\end{document}